\documentclass[aps,prd,amsmath,amssymb,preprintnumbers,12pt,twoside, a4paper]{article}
\def\pd{\partial}
\def\Slash{{\!\!\!\!/}}
\def\stru{\displaystyle\rule[-.8ex]{0ex}{2.9ex}}
\usepackage[dvipdfmx]{graphicx}
\usepackage{amsmath,amssymb,graphicx,epsfig,slashed}
\usepackage[hdivide={25mm,,25mm}, vdivide={25mm,,25mm}, nohead]{geometry}

\usepackage[normalem]{ulem} 

\usepackage{color}

\newcommand{\nn}{{\nonumber}}

\usepackage{subfigure}
\usepackage{fancyhdr}
\usepackage{enumitem}
\def\simlt{\mathrel{\lower2.5pt\vbox{\lineskip=0pt\baselineskip=0pt
             \hbox{$<$}\hbox{$\sim$}}}}
\def\simgt{\mathrel{\lower2.5pt\vbox{\lineskip=0pt\baselineskip=0pt
             \hbox{$>$}\hbox{$\sim$}}}}

\input{epsf.sty} 
 
\begin{document}

\thispagestyle{empty}

\begin{center}
{\Large{\textbf{Fayet-Iliopoulos terms in supergravity\\ and D-term inflation}}}
\\
\medskip
\vspace{1cm}
\textbf{
I.~Antoniadis$^{\,a,b,}$\footnote{antoniadis@itp.unibe.ch}, 
A.~Chatrabhuti$^{c,}$\footnote{dma3ac2@gmail.com}, 
H.~Isono$^{c,}$\footnote{hiroshi.isono81@gmail.com}, 
R.~Knoops$^{c,}$\footnote{rob.k@chula.ac.th}
}
\bigskip

$^a$ {\small Laboratoire de Physique Th\'eorique et Hautes Energies - LPTHE,\\ 
Sorbonne Universit\'e, CNRS, 4 Place Jussieu, 75005 Paris, France}

$^b$ {\small Albert Einstein Center, Institute for Theoretical Physics,
University of Bern,\\ Sidlerstrasse 5, CH-3012 Bern, Switzerland }

$^c$ {\small Department of Physics, Faculty of Science, Chulalongkorn University,
\\Phayathai Road, Pathumwan, Bangkok 10330, Thailand }

\end{center}

\vspace{1cm}

\begin{abstract}
We analyse the consequences of a new gauge invariant Fayet-Iliopoulos (FI) term proposed recently to a class of inflation models driven by supersymmetry breaking with the inflaton being the superpartner of the goldstino. We first show that charged matter fields can be consistently added with the new term, as well as the standard FI term in supergravity in a K\"ahler frame where the $U(1)$ is not an R-symmetry. We then show that the slow-roll conditions can be easily satisfied with inflation driven by a D-term depending on the two FI parameters. Inflation starts at initial conditions around the maximum of the potential where the $U(1)$ symmetry is restored and stops when the inflaton rolls down to the minimum describing the present phase of our Universe. The resulting tensor-to-scalar ratio of primordial perturbations can be even at observable values in the presence of higher order terms in the K\"ahler potential.

\end{abstract}

\newpage

\def\pd{\partial}
\def\Slash{{\!\!\!\!/}}
\def\stru{\displaystyle\rule[-.8ex]{0ex}{2.9ex}}

\section{Introduction}

In a recent work~\cite{previouspaper}, we proposed a class of minimal inflation models in supergravity that solve the $\eta$-problem in a natural way by identifying the inflaton with the goldstino superpartner in the presence of a gauged R-symmetry. The goldstino/inflaton superfield has then charge one, the superpotential is linear and the scalar potential has a maximum at the origin with a curvature fixed by the quartic correction to the K\"ahler potential $K$ expanded around the symmetric point. The D-term has a constant Fayet-Iliopoulos (FI) contribution but plays no role in inflation and can be neglected, while the pseudoscalar partner of the inflaton is absorbed by the $U(1)_R$ gauge field that becomes massive away from the origin.

Recently, a new FI term was proposed~\cite{Cribiori:2017laj} that has three important properties: (1) it is manifestly gauge invariant already at the Lagrangian level; (2) it is associated to a $U(1)$ that should not gauge an R-symmetry and (3) supersymmetry is broken by (at least) a D-auxiliary expectation value and the extra bosonic part of the action is reduced in the unitary gauge to a constant FI contribution leading to a positive shift of the scalar potential, in the absence of matter fields. In the presence of neutral matter fields, the FI contribution to the D-term acquires a special field dependence $e^{2K/3}$ that violates invariance under K\"ahler transformations.

In this work, we study the properties of the new FI term and explore its consequences to the class of inflation models we introduced in~\cite{previouspaper}.\footnote{
This new FI term was also studied in~\cite{Aldabergenov:2017hvp} to remove an instability from inflation in Polonyi-Starobinsky supergravity.} 
We first show that matter fields charged under the $U(1)$ gauge symmetry can consistently be added in the presence of the new FI term, as well as a non-trivial gauge kinetic function. 
We then observe that the new FI term is not invariant under K\"ahler transformations. On the other hand, a gauged R-symmetry in ordinary K\"ahler invariant supergravity can always be reduced to an ordinary (non-R) $U(1)$ by a K\"ahler transformation. By then going to such a frame, we find that the two FI contributions to the $U(1)$ D-term can coexist, leading to a novel contribution to the scalar potential.

The resulting D-term scalar potential provides an alternative realisation of inflation from supersymmetry breaking, driven by a D- instead of an F-term. The inflaton is still a superpartner of the goldstino which is now a gaugino within a massive vector multiplet, where again the pseudoscalar partner is absorbed by the gauge field away from the origin. For a particular choice of the inflaton charge, the scalar potential has a maximum at the origin where inflation occurs and a supersymmetric minimum at zero energy, in the limit of negligible F-term contribution (such as in the absence of superpotential). The slow roll conditions are automatically satisfied near the point where the new FI term cancels the charge of the inflaton, leading to higher than quadratic contributions due to its non trivial field dependence. 

The K\"ahler potential can be canonical, modulo the K\"ahler transformation that takes it to the non R-symmetry frame. In the presence of a small superpotential, the inflation is practically unchanged and driven by the D-term, as before. However, the maximum is now slightly shifted away from the origin and the minimum has a small non-vanishing positive vacuum energy, where supersymmetry is broken by both F- and D-auxiliary expectation values of similar magnitude. The model predicts in general small primordial gravitational waves with a tensor-to-scaler ration $r$ well below the observability limit. However, when higher order terms are included in the K\"ahler potential, one finds that $r$ can increase to large values $r\simeq 0.015$.

The outline of our paper is the following. 
In Section~\ref{newFI}, we review the new FI term (section 2.1)  
and we show that matter fields charged under the $U(1)$ gauge symmetry can consistently be added, as well as a non-trivial gauge kinetic function (section 2.2). We also find that besides the new FI term, the usual (constant) FI contribution to the D-term~\cite{Fayet:1974jb} can also be present. 
Next, we show that the new FI term breaks the K\"ahler invariance of the theory, and therefore forbids the presence of any gauged R-symmetries. As a result, the two FI terms can only coexist in the K\"ahler frame where the $U(1)$ is not an R-symmetry (section 2.3). 
In Section~\ref{nonRpotential}, we compute the resulting scalar potential and analyse its extrema and supersymmetry breaking in both cases of absence (section 3.1) or presence (section 3.2) of superpotential. 
In Section~\ref{applinflation}, we analyse the consequences of the new term in the models of inflation driven by supersymmetry breaking. We first consider a canonical K\"ahler potential (section 4.1) and then present a model predicting sizeable spectrum of primordial tensor fluctuations by introducing higher order corrections (section 4.2).

\section{On the new FI term}
\label{newFI}

In this section we follow the conventions of~\cite{Freedman:2012zz} and set the Planck mass to 1. 

\subsection{Review}

In~\cite{Cribiori:2017laj}, the authors propose a new contribution to the supergravity Lagrangian of the form\footnote{
A similar, but not identical term was studied in~\cite{Kuzenko:2018jlz}.}
\begin{align}
\mathcal L_{\text{FI}} &= \xi_2 \left[  S_0 \bar S_0 \frac{w^2 \bar w^2}{ \bar T(w^2)  T ( \bar w^2) } (V)_D \right]_D . \label{NewFI}
\end{align}
The chiral compensator field $S_0$, with Weyl and chiral weights $(\text{Weyl},\text{Chiral}) = (1,1)$, has components $S_0 = \left( s_0, P_L \Omega_0, F_0 \right)$ .
The vector multiplet has vanishing Weyl and chiral weights, and its components are given by $V = \left( v , \zeta, \mathcal H, v_\mu, \lambda, D \right)$.
In the Wess-Zumino gauge, the first components are put to zero $v = \zeta = \mathcal H = 0$.
The multiplet $w^2$ is of weights $(1,1)$, and given by
\begin{align}
 w^2 = \frac{ \bar \lambda P_L \lambda}{S_0^2}, \ \ \ \ \ \ 
 \bar w^2 = \frac{  \lambda P_R \bar \lambda}{\bar S_0^2}.
\end{align}
The components of $\bar \lambda P_L \lambda $ are given by
\begin{align}
\bar \lambda P_L \lambda = \Big( 
\bar \lambda P_L \lambda \ \ ; \ \
\sqrt{2} P_L \big( - \frac{1}{2} \gamma \cdot  \hat F + i D \big) \lambda \ \ ; \ \ 
2 \bar \lambda P_L \slashed{\cal D} \lambda + \hat F^- \cdot \hat F^- - D^2 \Big) . 
\end{align}

The kinetic terms for the gauge multiplet are given by
\begin{align}
 \mathcal L_{\text{kin}} = - \frac{1}{4} \left[ \bar \lambda P_L \lambda \right]_F + \text{h.c. }.
\end{align}
The operator $T$ ($\bar T$) is defined in~\cite{Kugo:1983mv,Ferrara:2016een}, 
and leads to a chiral (antichiral) multiplet. 
For example, the chiral multiplet $T(\bar w^2)$ has weights $(2,2)$. 
In global supersymmetry the operator $T$ corresponds to the usual chiral projection operator $\bar D^2$.\footnote{
The operator $T$ indeed has the property that $T(Z) = 0$ for a chiral multiplet $Z$. 
Moreover, for a vector multiplet $V$ we have $T(Z C) = Z T(C)$, and $[C]_D = \frac{1}{2} [T(C) ]_F$. }

From now on, we will drop the notation of $\text{h.c.}$ and implicitly assume its presence for every $[ \ \ ]_F$ term in the Lagrangian.
Finally, the multiplet $(V)_D$ is a linear multiplet with weights $(2,0)$, given by
\begin{align}
 (V)_D = \left( D, \slashed{\cal D} \lambda, 0 , {\cal D}^b \hat{F}_{ab}, - \slashed{\cal D} \slashed{\cal D} \lambda, - \Box^C {D}  
 \right) .
\end{align}
The definitions of $\slashed{\cal D} \lambda$ and the covariant field strength $\hat F_{ab}$ can be found in eq. (17.1) of \cite{Freedman:2012zz}, 
which reduce for an abelian gauge field to
\begin{align}
 \hat F_{ab} &= e_a^{\ \mu} e_b^{\ \nu} \left( 2 \partial_{[\mu} A_{\nu ]} + \bar \psi_{[\mu} \gamma_{\nu ]} \lambda \right) \notag \\
 {\cal D}_\mu \lambda &= \left( \partial_\mu - \frac{3}{2} b_\mu + \frac{1}{4} w_\mu^{ab} \gamma_{ab} - \frac{3}{2} i \gamma_* \mathcal A_\mu \right) \lambda
 -\left( \frac{1}{4} \gamma^{ab} \hat F_{ab} + \frac{1}{2} i \gamma_* D \right) \psi_\mu .
\end{align}
Here, $e_a^{\ \mu}$ is the vierbein, with frame indices $a,b$ and coordinate indices $\mu, \nu$.
The fields $w_\mu^{ab}$, $b_\mu$, and $\mathcal A_\mu$ are the gauge fields corresponding to Lorentz transformations, dilatations, and $T_R$ symmetry of the conformal algebra respectively,
while $\psi_\mu$ is the gravitino. The conformal d'Alembertian is given by $\Box^C = \eta^{ab} {\cal D}_a {\cal D}_b $.

It is important to note that the FI term given by eq.~(\ref{NewFI}) does not require the gauging of an R-symmetry, but breaks invariance under K\"ahler transformations. 
In fact, we will show below that a gauged R-symmetry would forbid such a term $\mathcal L_{FI}$.\footnote{We kept the notation of~\cite{Cribiori:2017laj}. 
Note that in this notation the field strength superfield $\mathcal W_\alpha$ is given by $\mathcal W ^2 = \bar \lambda P_L \lambda$, 
and $(V)_D$ corresponds to $\cal D^\alpha \mathcal W_\alpha$.}

The resulting Lagrangian after integrating out the auxiliary field $D$ contains a term 
\begin{align}
 \mathcal L_{\text{FI,new}} = - \frac{ \xi_2^2}{2}  \left( s_0 \bar s_0 \right)^2 . \label{onlyFI_VD}
\end{align}
In the absence of additional matter fields, one can use the Poincar\'e gauge $s_0 = \bar s_0 = 1$,
resulting in a constant D-term contribution to the scalar potential. 
This prefactor however is relevant when matter couplings are included in the next section.

\subsection{Adding (charged) matter fields}

In this section we couple the term $\mathcal L_{\text{FI}}$ given by eq.~(\ref{NewFI}) to additional matter fields charged under the $U(1)$.
For simplicity, we focus on a single chiral multiplet $X$. 
The extension to more chiral multiplets is trivial.
The Lagrangian is given by
\begin{align}
 \mathcal L = -3 \left[ S_0 \bar S_0 e^{- \frac{1}{3} K(X ,\bar X) } \right]_D + \left[ S_0^3 W(X) \right]_F 
 -\frac{1}{4} \left[ f(X) \bar \lambda P_L \lambda \right]_F  +  \mathcal L_{\text{FI}} , \label{Full_Lagrangian}
\end{align}
with a K\"ahler potential $K(X, \bar X)$, a superpotential $W(X)$ and a gauge kinetic function $f(X)$. 
The first three terms in eq.~(\ref{Full_Lagrangian}) give the usual supergravity Lagrangian 
\cite{Freedman:2012zz}. 
We assume that the multiplet $X$ transforms under the $U(1)$,
\begin{align}
 V &\rightarrow V + \Lambda + \bar \Lambda, \notag \\
 X &\rightarrow X e^{- q \Lambda}, \label{XVgaugetransf}
\end{align}
with gauge multiplet parameter $\Lambda$.
We assume that the $U(1)$ is not an R-symmetry. In other words, we assume that the superpotential does not transform under the gauge symmetry.
The reason for this will be discussed in section~\ref{sec_kahlerR}.
For a model with a single chiral multiplet this implies that the superpotential is constant
\begin{align}
 W(X) = F . \label{constantsuperpotential}
\end{align}
Gauge invariance fixes the K\"ahler potential to be a function of $X e^{qV} \bar X$ (for notational simplicity, in the following we omit the $e^{qV}$ factors). 

Indeed, in this case the term $\mathcal L_{\text{FI}}$ can be consistently added to the theory, similar to~\cite{Cribiori:2017laj},
and the resulting D-term contribution to the scalar potential acquires an extra term proportional to $\xi_2$
\begin{align}
 \mathcal V_D &= \frac{1}{2} \text{Re}\left( f(X) \right)^{-1}   \left( i k_X \partial_X K + \xi_2 e^{\frac{2}{3} K}  \right)^2 , \label{VDextra}
\end{align}
where the Killing vector is $k_X = -iqX$ and $f(X)$ is the gauge kinetic function.
The F-term contribution to the scalar potential remains the usual
\begin{align}
 \mathcal V_F &= e^{K(X,\bar X)} \left( -3 W \bar W + g^{X \bar X} \nabla_X W \bar \nabla_{\bar X} \bar W \right) .
\end{align}
For a constant superpotential (\ref{constantsuperpotential}) this reduces to
\begin{align}
 \mathcal V_F &= |F|^2 e^{K(X, \bar X)} \left( -3 + g^{X \bar X} \partial_X K \partial_{\bar X} K \right) .
\end{align}

From eq.~(\ref{VDextra}) it can be seen that if the K\"ahler potential includes a term proportional to $\xi_1 \log(X \bar X)$, 
the D-term contribution to the scalar potential acquires another constant contribution. 
For example, if 
\begin{align}
 K (X , \bar X) = X \bar X + \xi_1 \ln(X \bar X), \label{Kahlerpotential}
\end{align}
the D-term contribution to the scalar potential becomes 
\begin{align}
 \mathcal V_D &= \frac{1}{2} \text{Re}\left( f(X) \right)^{-1}   \left( q X \bar X + q \xi_1 + \xi_2 e^{\frac{2}{3} K}  \right)^2 .
\end{align}
We will argue below that the contribution proportional to $\xi_1$ is the usual FI term in a non R-symmetric K\"ahler frame, which can be consistently added to the model including the new FI term proportional to $\xi_2$.

In the absence of the extra term, a K\"ahler transformation
\begin{align}
 K(X , \bar X) &\rightarrow K(X ,\bar X)  + J(X) + \bar J(\bar X), \notag \\
 W(X) &\rightarrow W(X) e^{-J(X)}, \label{kahlertransformation}
\end{align}
with $J(X) = - \xi_1 \ln X$ allows one to recast the model in the form
\begin{align}
 K(X, \bar X) &= X \bar X , \notag \\
 W(X) &= m_{3/2} X . \label{Kahlertransformedmodel}
\end{align}
The two models result in the same Lagrangian, at least classically\footnote{
At the quantum level, a K\"ahler transformation also introduces a change in the gauge kinetic function $f$, see for example~\cite{Kaplunovsky:1994fg}.}.
However, in the K\"ahler frame of eqs.~(\ref{Kahlertransformedmodel}) the superpotential transforms nontrivially under the gauge symmetry.
As a consequence, the gauge symmetry becomes an R-symmetry. 
We will argue in section~\ref{sec_kahlerR} that 
\begin{enumerate}
 \item The extra term~(\ref{NewFI}) violates the K\"ahler invariance of the theory, and the two models related by a K\"ahler transformation are no longer equivalent.
 \item The model written in the K\"ahler frame where the gauge symmetry becomes an R-symmetry in eqs.~(\ref{Kahlertransformedmodel}) can not be consistently coupled to $\mathcal L_{\text{FI}}$.
\end{enumerate}

\subsection{K\"ahler invariance and R-symmetry} \label{sec_kahlerR}

In the superconformal compensator formalism of supergravity the chiral compensator $S_0$ is not uniquely defined:
a redefinition of the chiral compensator $S_0' = S_0 e^{J/3}$ results in a K\"ahler transformation (\ref{kahlertransformation}) with parameter $J$. 
In other words, the chiral compensator field $S_0$ transforms under a K\"ahler transformation as
\begin{align}
 K(X , \bar X) &\rightarrow K(X ,\bar X)  + J(X) + \bar J(\bar X), \notag \\
 W(X) &\rightarrow W(X) e^{-J(X)} , \notag \\
  S_0 &\rightarrow  S_0 e^{\frac{J(X)}{3}} \label{chiralKahler} .
\end{align}
Indeed, the first three terms in the Lagrangian~(\ref{Full_Lagrangian}) are invariant under this transformation. 
However, $\mathcal L_{\text{FI}}$, repeated here for convenience,
\begin{align}
\mathcal L_{\text{FI}} &= \xi_2 \left[  S_0 \bar S_0 \frac{w^2 \bar w^2}{ \bar T(w^2)  T ( \bar w^2) } (V)_D \right]_D , \notag
\end{align}
is not invariant under eqs.~(\ref{chiralKahler}). 
In particular,
\begin{align}
 w^2 &\rightarrow w^2 e^{-\frac{2}{3} J(X)} ,   		&	\bar w^2 &\rightarrow \bar w^2 e^{-\frac{2}{3} \bar J(X)},  \notag \\
 \bar T (w^2) &\rightarrow \bar T( w^2 e^{-\frac{2}{3} J(X)} ),	&	T( \bar w^2)  &\rightarrow T ( \bar w^2 e^{-\frac{2}{3} \bar J(X)} ) ,
\end{align}
where the first component of $\bar T (w^2  e^{-\frac{2}{3} J(X)}) $ is given by (up to 4-fermion terms)
\begin{align}
  \bar T (   w^2 e^{-\frac{2}{3}  J} ) = \Big(
 & \frac{e^{-2j/3}}{s_0^2} \Big[ 
  2 \bar \lambda P_L \slashed{\cal D} \lambda + F^- \cdot F^{-} - D^2
 - 2 \bar \lambda P_L \lambda \left( \frac{1}{s_0} F_0 + \frac{1}{3} F_j \right)  \\
 & + \frac{2 \sqrt{2}}{s_0} \left( \bar \chi_0 + \frac{1}{3} \bar \chi_j \right) P_L \left( -\frac{1}{2} \gamma \cdot F + i D \right) \lambda
 \Big]  + \text{4-fermions} \ ; \dots \ ; \ \dots \Big), \notag 
\end{align}
and we used the notation $J(X) = ( j, P_L \chi_j,  F_j)$.
As a comparison, the lowest component of $\bar T(  w^2 ) e^{-\frac{2}{3}  J} $ does not contain $\chi_j$ and $F_j$, and is given by
\begin{align}
 \bar T ( w^2 ) e^{-\frac{2}{3}J} =   \Big(
& \frac{e^{-2j/3}}{s_0^2} \Big[ 
  2 \bar \lambda P_L \slashed{\cal D} \lambda + F^- \cdot F^{-} - D^2
 -  \frac{2}{s_0} \bar \lambda P_L \lambda   F_0   \\
 & + \frac{2 \sqrt{2}}{s_0}  \bar \chi_0  P_L \left( -\frac{1}{2} \gamma \cdot F + i D \right) \lambda
 \Big]  + \text{4-fermions} \ ; \dots \ ; \ \dots \Big). \notag 
\end{align}
As a result $\bar T(w^2)$ is not K\"ahler covariant, and the term $\mathcal L_{\text{FI}}$ violates the K\"ahler invariance of the theory.\footnote{
It would be interesting to see if this term can be modified in order to keep K\"ahler invariance 
without modifying the resulting contribution to the scalar potential eq.~(\ref{onlyFI_VD}), at least in the rigid limit~\cite{nextproject}.}

As a result, a K\"ahler transformation is no longer a symmetry of the Lagrangian.
Therefore, when the term $\mathcal L_{\text{FI}}$ is included, two models that are related by a K\"ahler transformation are no longer (classically) equivalent.
Nevertheless, it is still possible to add the usual FI term $\xi_1$ in the K\"ahler frame where the gauge symmetry is not an R-symmetry. 

Indeed, a $U(1)_R$-symmetry is a $U(1)$ gauge symmetry under which the superpotential transforms with a phase. 
For example, under the gauge transformations eq.~(\ref{XVgaugetransf}) the superpotential in eq.~(\ref{Kahlertransformedmodel}) transforms as
\begin{align}
 W \rightarrow W e^{- q \Lambda},
\end{align}
and the chiral compensator field transforms as\footnote{
Note that this is technically not yet an R-symmetry: after fixing the conformal gauge, a mixture of the $U(1)_R$ described above and the 
T-symmetry in the superconformal algebra is broken down to the usual R-symmetry in supergravity $U(1)_R'$:
$U(1)_R \times U(1)_T \rightarrow U(1)_R'$.}
\begin{align}
 S_0 \rightarrow S_0 e^{ \frac{q}{3} \Lambda}. \label{compensatorU1}
\end{align}
Following the same arguments as above for the K\"ahler transformations, one can see that the gauge invariance is violated in $\mathcal L_{\text{FI}}$ as a consequence of eq.~(\ref{compensatorU1}). 
As a result, in order for the extra FI term to be consistently coupled to supergravity, the superpotential should not transform. 
Although the usual FI term in supergravity is usually associated with a gauged R-symmetry~\cite{Freedman:1976uk,Barbieri:1982ac},
it is possible to rewrite a theory with a non-zero superpotential (by a K\"ahler transformation) in terms of a constant superpotential 
and the K\"ahler potential 
given by $K' = K + \ln(W \bar W)$. This leads to the same (classical) Lagrangian, while the gauge symmetry that gave rise to the constant FI contribution is not an R-symmetry. 
Therefore, in this 'K\"ahler frame', the theory can be consistently coupled to $\mathcal L_{\text{FI}}$. 
In fact, this was the motivation behind the choice of the K\"ahler potential in eq.~(\ref{Kahlerpotential}).

\section{The scalar potential in a Non R-symmetry frame}
\label{nonRpotential}

In this section, we work in the K\"ahler frame where the superpotential does not transform, 
and take into account the two types of FI terms which were discussed in the last section.
For convenience, we repeat here the K\"ahler potential in eq.~(\ref{Kahlerpotential}) 
and restore the inverse reduced Planck mass $\kappa=M_{\rm Pl}^{-1}=(2.4 \times 10^{18} \, {\rm GeV})^{-1}$: 
\begin{equation}
K = \kappa^{-2} (X\bar{X} + \xi_1 \ln X\bar{X}).
\label{Kahlerpotential2}
\end{equation}
The superpotential and the gauge kinetic function are set to be constant\footnote{
Strictly speaking, the gauge kinetic function gets a field-dependent correction proportional to $q^2\ln \rho$, 
in order to cancel the chiral anomalies~\cite{previouspaper}. 
However, the correction turns out to be very small and can be neglected
below, since the charge $q$ is chosen to be of order of $10^{-5}$ or smaller.
}:
\begin{equation}
W = \kappa^{-3} F  , \qquad f(X) = 1.
\label{const_W}
\end{equation}
After performing a change of the field variable $X = \rho e^{i\theta}$ where $\rho \geq 0$ and setting $\xi_1 = b$, the full scalar potential $\mathcal{V} = \mathcal{V}_F + \mathcal{V}_D$ is a function of $\rho$. The F-term contribution to the scalar potential is given by
\begin{equation}
\label{VF}
\mathcal{V}_F = \frac{1}{\kappa ^4}F^2 e^{\rho ^2} \rho ^{2 b} \left[\frac{\left(b+\rho ^2\right)^2}{\rho ^2}-3\right],
\end{equation}
and the D-term contribution is 
\begin{equation}
\label{VDxi2}
\mathcal{V}_D = \frac{q^2}{2 \kappa ^4} \left(b+\rho ^2+\xi   \rho ^{\frac{4 b}{3}}e^{\frac{2}{3} \rho ^2}\right)^2.
\end{equation}
Note that we rescaled the second FI parameter by $\xi = \xi_2/q$.  
We consider the case with $\xi \neq 0$ because we are interested in the role of the new FI-term in inflationary models driven by supersymmetry breaking. 
{Moreover, the limit $\xi \rightarrow 0$ is ill-defined~\cite{Cribiori:2017laj}.}

The first FI parameter $b$ was introduced as a free parameter. We now proceed to narrowing the value of $b$ by the following physical requirements. 
We first consider the behaviour of the potential around $\rho=0$,
\begin{align}
\mathcal{V}_D &=\frac{q^2}{2 \kappa ^4} 
\Big[\big(b^2+2b\rho^2+O(\rho^4)\big)
+ 2b\xi\rho^{\frac{4b}{3}}\big(1+O(\rho^2)\big)
+ \xi^2\rho^{\frac{8b}{3}}\big(1+O(\rho^2)\big)
\Big], \\
\mathcal{V}_F &= \frac{F^2}{\kappa ^4} \rho^{2b}\Big[b^2\rho^{-2}+(2b-3)+O(\rho^2)\Big].
\end{align} 
In this paper we are interested in small-field inflation models in which the inflation starts in the neighbourhood of a local maximum at $\rho=0$. 
In our last paper \cite{previouspaper}, we considered models of this type with $\xi=0$ (which were called Case 1 models), and found that the choice $b=1$ is forced by the requirement that the potential takes a finite value at the local maximum $\rho=0$. In this paper, we will investigate the effect of the new FI parameter $\xi$ on the choice of $b$ under the same requirement.

First, in order for $\mathcal{V}(0)$ to be finite, we need $b \geq 0$.
We first consider the case $b>0$. 
We next investigate the condition that the potential at $\rho=0$ has a local maximum.
For clarity we discuss below the cases of $F=0$ and $F \neq 0 $ separately.
The $b=0$ case will be treated at the end of this section.

\subsection{Case $F=0$}
\label{F0}

In this case $\mathcal V_F = 0$ and the scalar potential is given by only the D-term contribution $\mathcal V = \mathcal V_D$.
Let us first discuss the first derivative of the potential:
\begin{align}
\mathcal{V}_D' = \frac{q^2}{2 \kappa^4}
\bigg[
4b\rho\big(1+O(\rho^2)\big)
+ \frac{8b^2}{3}\xi\rho^{\frac{4b}{3}-1}\big(1+O(\rho^2)\big)
+ \frac{8b}{3}\xi^2\rho^{\frac{8b}{3}-1}\big(1+O(\rho^2)\big)
\bigg].
\end{align}
For $\mathcal{V}_D'(0)$ to be convergent, we need $b \geq 3/4$ (note that $\xi \neq 0$).
When $b=3/4$, we have $\mathcal{V}_D'(0)=8b^2\xi/3$,
which does not give an extremum because we chose $\xi \neq 0$.
On the other hand, when $b>3/4$, we have $\mathcal{V}_D'(0)=0$.
To narrow the allowed value of $b$ further, let us turn to the second derivative,
\begin{align}
\mathcal{V}_D'' &= \frac{q^2}{2\kappa^4}
\bigg[
4b\big(1+O(\rho^2)\big) + \frac{8b^2}{3}\Big(\frac{4b}{3}-1\Big)\xi\rho^{\frac{4b}{3}-2}\big(1+O(\rho^2)\big)
\nn\\
&{} \qquad\qquad\qquad
+ \frac{8b}{3}\Big(\frac{8b}{3}-1\Big)\xi^2\rho^{\frac{8b}{3}-2}\big(1+O(\rho^2)\big)
\bigg].
\end{align}
When $3/4<b<3/2$, the second derivative $\mathcal{V}_D''(0)$ diverges. 
When $b>3/2$, the second derivative becomes $\mathcal{V}_D''(0)=2\kappa^{-4} q^2b>0$, which gives a minimum. 

We therefore conclude that to have a local maximum at $\rho=0$, we need to choose $b=3/2$, for which we have
\begin{align}
\mathcal{V}_D''(0) = 3\kappa^{-4} q^2(\xi+1).
\end{align}
The condition that $\rho =0 $ is  a local maximum requires $\xi<-1$.

Let us next discuss the global minimum of the potential with $b=3/2$ and $\xi<-1$.
The first derivative of the potential without approximation reads
\begin{align}
\mathcal{V}_D' \propto \rho(3+3\xi e^{\frac{2}{3}\rho^2}+ 2 \xi\rho^2 e^{\frac{2}{3}\rho^2})
(3+2\rho^2+2\xi\rho^2 e^{\frac{2}{3}\rho^2}).
\end{align}
Since $3+3\xi e^{\frac{2}{3}\rho^2}+ 2 \xi\rho^2 e^{\frac{2}{3}\rho^2}<0$ for $\rho \geq 0$ and $\xi<-1$, the extremum away from $\rho=0$ is located at $\rho_v$ satisfying the condition
\begin{align}
3+2\rho_v^2+2\xi\rho_v^2 e^{\frac{2}{3}\rho_v^2} = 0.
\end{align}
Substituting this condition into the potential $\mathcal{V}_D$ gives $\mathcal{V}_D(\rho_v)=0$.

We conclude that for $\xi < -1$ and $b= 3/2$ the potential has a maximum at $\rho = 0$, and a supersymmetric minimum at $\rho_v$. 
We postpone the analysis of inflation near the maximum of the potential in section~\ref{applinflation}, and  
the discussion of the uplifting of the minimum in order to obtain a small but positive cosmological constant below. 
In the next subsection we investigate the case $F \neq 0$.

We finally comment on supersymmetry (SUSY) breaking in the scalar potential.
Since the superpotential is zero, the SUSY breaking is measured by the D-term order parameter, namely the Killing potential associated with the gauged ${\rm U}(1)$, which is defined by
\begin{align}
\mathcal{D} &= i\kappa^{-2}\frac{-iqX}{ W}\bigg(\frac{\pd  W}{\pd X}
+\kappa^2 \frac{\pd \mathcal K}{\pd X} W\bigg).
\end{align}
This enters the scalar potential as $\mathcal{V}_D=\mathcal{D}^2/2$.  So, at the local maximum and during inflation $\mathcal{D}$ is of order $q$ and supersymmetry is broken. On the other hand, at the global minimum,  supersymmetry is preserved and the potential vanishes.

\subsection{Case $F \neq 0$}
\label{Fnot0}

In this section we take into account the effect of $\mathcal{V}_F$; its first derivative reads:
\begin{align}
\mathcal{V}_F' &= \kappa^{-4}F^2 
\bigg[
b^2(2b-2)\rho^{2b-3} + 2b(2b-3)\rho^{2b-1}\big(1+O(\rho^2)\big)
\bigg]\, .
\end{align}
For $\mathcal{V}'(0)$ to be convergent, we need $b \geq 3/2$, for which $\mathcal{V}_D'(0)=0$ holds.
For $b=3/2$, we have $\mathcal{V}_F'(0)=(9/4)\kappa^{-4}F^2>0$, that does not give an extremum.
For $b>3/2$, we have $\mathcal{V}_F'(0)=0$. To narrow the allowed values of $b$ further, let us turn to the second derivative,
\begin{align}
\mathcal{V}_F'' &= \kappa^{-4}F^2 \bigg[
b^2(2b-2)(2b-3)\rho^{2b-4} + 2b(2b-3)(2b-1)\rho^{2b-2}\big(1+O(\rho^2)\big)
\bigg]\, .
\end{align}
For $3/2<b<2$, the second derivative $\mathcal{V}_F''(0)$ diverges.
For $b \geq 2$, the second derivative is positive $\mathcal{V}''(0)>0$, that gives a minimum (note that $V_D''(0) > 0$ as well in this range).

We conclude that the potential cannot have a local maximum at $\rho=0$ for any choice of $b$.
Nevertheless, as we will show below, the potential can have a local maximum in the neighbourhood of $\rho=0$ if we choose $b=3/2$ and $\xi < -1$.
For this choice, the derivatives of the potential have the following properties,
\begin{align}
\mathcal{V}'(0)<0, \quad \mathcal{V}''(0) = 3\kappa^{-4} q^2(\xi+1).
\end{align}
The extremisation condition around $\rho=0$ becomes
\begin{align}
3\kappa^{-4} q^2 (\xi+1) \rho + \frac{9}{4}\kappa^{-4}F^2 \simeq 0 .
\end{align}
So the extremum is at
\begin{align}
\rho \simeq -\frac{3F^2}{4q^2(\xi+1)} .
\end{align}
Note that the extremum is in the neighbourhood of $\rho=0$ 
as long as we keep the $F$-contribution to the scalar potential small
by taking $F^2 \ll q^2|\xi+1|$, which guarantees the approximation ignoring higher order terms in $\rho$. We now choose $\xi<-1$ so that $\rho$ for this extremum is positive.
The second derivative at the extremum reads
\begin{align}
\mathcal{V}'' \simeq 3\kappa^{-4} q^2(\xi+1) ,
\end{align}
as long as we ignore higher order terms in $F^2/(q^2|\xi+1|)$.
By our choice $\xi<-1$, the extremum is a local maximum, as desired.

\begin{figure*}
\begin{center}
  \includegraphics[width=0.5\linewidth]{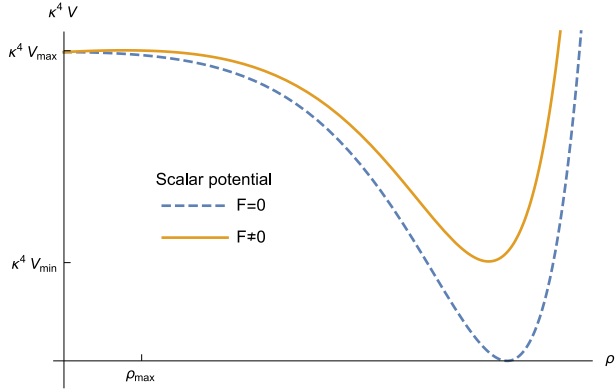}
 \caption{\small This plot shows the scalar potentials in $F=0$ and $F\neq 0$ cases. When $F=0$, we have a local maximum at $\rho_{\rm max} = 0$ and a global minimum with zero cosmological constant.  For $F \neq 0$, the local maximum is shifted by a small positive value to $\rho_{\rm max} \neq 0$. The global minimum now has a positive cosmological constant.}
 \label{potential}
\end{center}
\end{figure*}

Let us comment on the global minimum after turning on the F-term contribution.
As long as we choose the parameters so that $F^2/q^2 \ll 1$, the change in the global minimum $\rho_v$ is very small, of order $\mathcal{O}(F^2/q^2)$, because the extremisation condition depends only on the ratio $F^2/q^2$. So the change in the value of the global minimum is of order $\mathcal{O}(F^2)$.
The plot of this change is given in Fig.~\ref{potential}.

In the present case $F \neq 0$, the order parameters of SUSY breaking are both the Killing potential $\mathcal{D}$ and the F-term contribution $\mathcal{F}_X$, which read
\begin{align}
\mathcal{D} &\propto q(\tfrac{3}{2}+\rho^2), \quad
\mathcal{F}_X \propto F\rho^{1/2} e^{\rho^2/2},
\end{align}
where the F-term order parameter $\mathcal{F}_X$ is defined by
\begin{align}
\mathcal{F}_X &= -\frac{1}{\sqrt{2}} e^{\kappa^2 \mathcal K/2} 
\bigg(\frac{\pd^2 \mathcal K}{\pd X\pd\bar X}\bigg)^{-1}
\bigg(\frac{\pd\bar{ W}}{\pd\bar X}+\kappa^2 \frac{\pd \mathcal K}{\pd\bar X}\bar{ W} \bigg).
\end{align}
Therefore, at the local maximum, $\mathcal{F}_X/\mathcal{D}$ is of order $\mathcal{O}((\xi+1)^{-1/2}F^2/q^2)$ because $\rho$ there is of order $\mathcal{O}((\xi+1)^{-1}F^2/q^2)$.
On the other hand, at the global minimum, both $\mathcal{D}$ and $\mathcal{F}_X$ are of order $\mathcal{O}(F)$, assuming that $\rho$ at the minimum is of order $\mathcal{O}(1)$, which is true in our models below. 
This makes tuning of the vacuum energy between the F- and D-contribution in principle possible, along the lines of~\cite{previouspaper,Antoniadis:2016aal}.

A comment must be made here on the action in the presence of non-vanishing $F$ and $\xi$. As mentioned above, the supersymmetry is broken both by the gauge sector and by the matter sector. The associated goldstino therefore consists of a linear combination of the $U(1)$ gaugino and the fermion in the matter chiral multiplet $X$. In the unitary gauge the goldstino is set to zero, so the gaugino  is not vanishing anymore, and the action does not simplify as in Ref.~\cite{Cribiori:2017laj}. This, however, only affects the part of the action with fermions, while the scalar potential does not change. This is why we nevertheless used the scalar potential \eqref{VF} and \eqref{VDxi2}. 

Let us consider now the case $b=0$ where only the new FI parameter $\xi$ contributes to the potential. In this case, the condition for the local maximum of the scalar potential at $\rho = 0$ can be satisfied for $-\frac{3}{2} < \xi < 0$.
When $F$ is set to zero, the scalar potential (\ref{VDxi2}) has a minimum at $\rho_{\text{min}}^2 = \frac{3}{2}\ln\big(-\frac{3}{2\xi} \big)$.  In order to have $\mathcal{V}_{\text{min}} = 0$, we can choose $\xi = - \frac{3}{2 e}$.  However, we find that this choice of parameter $\xi$ does not allow slow-roll inflation near the maximum of the scalar potential.   Similar to our previous models \cite{previouspaper}, it may be possible to achieve both the scalar potential satisfying slow-roll conditions and a small cosmological constant at the minimum by adding correction terms to the K\"ahler potential and turning on a parameter $F$.  However, in this paper, we will focus on $b = 3/2$ case where, as we will see shortly, less parameters are required to satisfy the observational constraints.

\section{Application in Inflation}
\label{applinflation}

In the previous work~\cite{previouspaper}, 
we proposed a class of supergravity models for small field inflation in which the inflation is identified with the sgoldstino, 
carrying a $U(1)$ charge under a gauged R-symmetry. In these models, inflation occurs around the maximum of the scalar potential, where the $U(1)$ symmetry is restored, with the inflaton rolling down towards the electroweak minimum.
These models also avoid the so-called $\eta$-problem in supergravity by taking a linear superpotential, $W \propto X$. 
In contrast, in the present paper we construct models with two FI parameters $b,\xi$ in the K\"ahler frame where the $U(1)$ gauge symmetry is not an R-symmetry. 
If the new FI term $\xi$ is zero, our models are K\"ahler equivalent to those with a linear superpotential in \cite{previouspaper} (Case 1 models with $b=1$).
The presence of non-vanishing $\xi$, however, breaks the K\"ahler invariance as shown in Section~\ref{newFI}. Moreover, the FI parameter $b$ cannot be 1 but is forced to be $b=3/2$, according to the argument in Section~\ref{nonRpotential}. So the new models do not seem to avoid the $\eta$-problem. Nevertheless, we will show below that this is not the case and the new models with $b = 3/2$ avoid the $\eta$-problem thanks to the other FI parameter $\xi$ which is chosen near the value at which the effective charge of $X$ vanishes between the two FI-terms. Inflation is again driven from supersymmetry breaking but from a D-term rather than an F-term as we had before.

\subsection{Example for slow-roll D-term inflation}

In this section we focus on the case where $b = 3/2$ and derive the condition that leads to slow-roll inflation scenarios, where the start of inflation (or, horizon crossing) is near the maximum of the potential at $\rho =0$.   We also assume that the scalar potential is D-term dominated by choosing $F =0$, for which the model has only two parameters, namely $q$ and $\xi$.   The parameter $q$ controls the overall scale of the potential and it will be fixed by the amplitude $A_s$ of the CMB data.  The only free-parameter left over is $\xi$, which can be tuned to satisfy the slow-roll condition.

In order to calculate the slow-roll parameters, we need to work with the canonically normalised field $\chi$ defined by
\begin{align}
\frac{d\chi}{d \rho} = \sqrt{2 g_{X\bar{X}}}.
\end{align}
The slow-roll parameters are given in terms of the canonical field $\chi$ by
\begin{align}
\epsilon = \frac{1}{2\kappa^2} \left( \frac{dV/d\chi}{V}\right)^2, \quad
 \eta = \frac{1}{\kappa^2} \frac{d^2V/d\chi^2}{V}. 
 \label{slowroll_pars1}
\end{align}
Since we assume inflation to start near $\rho = 0$, the slow-roll parameters for small $\rho$ can be expanded as 
\begin{align}
 \epsilon &=  \frac{F^4}{q^4}+\frac{4 F^2  \left(2 (\xi +1) q^4-3 F^4\right)}{3 q^6}\rho\notag\\
& +\left(\frac{16}{9} (\xi +1)^2+\frac{2 F^4 \left(18 F^4-q^4 (20 \xi +11)\right)}{3 q^8}\right) \rho ^2   + \mathcal O(\rho^3),  \notag \\
 \eta &= \frac{4(1+ \xi)}{3} + \mathcal{O}(\rho) \label{slowroll_expand} .
\end{align}
Note also that $\eta$ is negative when $\xi<-1$.  We can therefore tune the parameter $\xi$ to avoid the $\eta$-problem.  
The observation is that at $\xi=-1$, the effective charge of $X$ vanishes and thus the $\rho$-dependence in the D-term contribution (\ref{VDxi2}) becomes of quartic order.

For our present choice $F=0$, the potential and the slow-roll parameters become functions of $\rho^2$ and the slow-roll parameters for small $\rho^2$ read
\begin{align}
 \eta &= \frac{4(1+ \xi)}{3} + \mathcal{O}(\rho^2)\, ,  \notag \\
 \epsilon &=  \frac{16}{9} (\xi +1)^2 \rho ^2   + \mathcal O(\rho^4)
 \simeq \eta(0)^2\rho^2 \, . \label{slowroll_expandF0}
\end{align}
Note that we obtain the same relation between $\epsilon$ and $\eta$ as in the model of inflation from supersymmetry breaking driven by an F-term from a linear superpotential and $b=1$~\cite{previouspaper}.
Thus, there is a possibility to have flat plateau near the maximum that satisfies the slow-roll condition and at the same time a small cosmological constant at the minimum nearby.  

 The number of e-folds $N$ during inflation is determined by
 \begin{align}
 N = \kappa^2\int^{\chi_{\rm end}}_{\chi_{*}}\frac{\mathcal{V}}{\partial_{\chi}\mathcal{V}} d \chi = \kappa^2\int^{\rho_{\rm end}}_{\rho_{*}}\frac{\mathcal{V}}{\partial_{\rho}\mathcal{V}}\left(\frac{d\chi}{d\rho} \right)^2 d \rho,
 \end{align}
where we choose $|\epsilon(\chi_{\rm end})| =1$.  
Notice that the slow-roll parameters for small $\rho^2$ satisfy the simple relation $\epsilon=\eta(0)^2\rho^2+O(\rho^4)$ by eq.~\eqref{slowroll_expandF0}. Therefore, the number of e-folds between $\rho=\rho_1$ and $\rho_2$ ($\rho_1<\rho_2$) takes the following simple approximate form as in \cite{previouspaper},
\begin{align}
\label{simpleNe}
N \simeq \frac{1}{|\eta(0)|} \ln \left( \frac{\rho_2}{\rho_1} \right)
= \frac{3}{4|\xi+1|} \ln \left( \frac{\rho_2}{\rho_1} \right).
\end{align}
as long as the expansions in \eqref{slowroll_expandF0} are valid in the region $\rho_1 \leq \rho \leq \rho_2$. Here we also used the approximation $\eta(0) \simeq \eta_*$, which holds in this approximation.

We can compare the theoretical predictions of our model to the observational data via the power spectrum of scalar perturbations of the CMB, namely the amplitude $A_s$, tilt $n_s$ and the tensor-to-scalar ratio of primordial fluctuations $r$.   These are written in terms of the slow-roll parameters:
\begin{align}
 A_s &=  \frac{\kappa^4 \mathcal{V}_*}{24\pi^2\epsilon_*}\, ,  \notag\\
 n_s &= 1 + 2\eta_* - 6\epsilon_*\simeq 1+2\eta_*         \, , \notag \\
 r &= 16\epsilon_*\, , 
 \end{align}
where all parameters are evaluated at the field value at horizon crossing $\chi_*$. From the relation of  the spectral index above, one should have $\eta_*\simeq -0.02$,  and thus eq.~(\ref{simpleNe}) gives approximately the desired number of e-folds when the logarithm is of order one.
Actually, using this formula, we can estimate the upper bound of the tensor-to-scalar ratio $r$ and the Hubble scale $H_*$ following the same argument given in \cite{previouspaper}; that is, the upper bounds are given by computing the parameters $r,H_*$ assuming that the expansions \eqref{slowroll_expandF0} hold until the end of inflation. We then get the bound
\begin{align}
r \lesssim 16(|\eta_*|\rho_{\rm end} e^{-|\eta_*|N})^2 \simeq 10^{-4}, \quad H_* \lesssim 10^{12} \, {\rm GeV},
\end{align}
where we used $|\eta_*|=0.02$, $N \simeq 50\,-\,60$ and $\rho_{\rm end} \lesssim 0.5$, which are consistent with our models. In the next subsection, we will present a model which gives a tensor-to-scalar ratio bigger than the upper bound above, by adding some perturbative corrections to the K\"ahler potential.

\begin{figure*}
\begin{center}
  \includegraphics[width=0.5\linewidth]{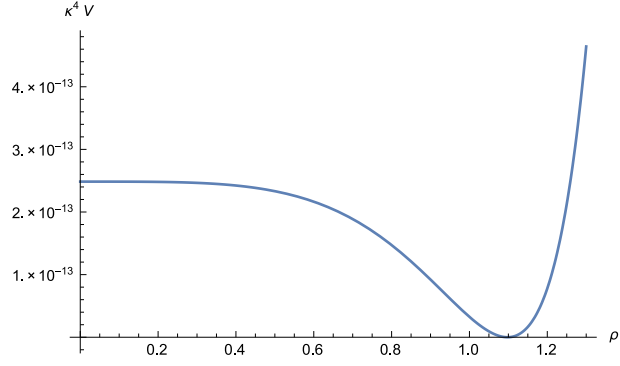}
 \caption{\small This plot shows the scalar potential for  $F =0$, $b=3/2$, $\xi = -1.005$ and $q = 4.544 \times 10^{-7}$.}
 \label{potential_AB_Zero}
\end{center}
\end{figure*}

\begin{figure*}
\begin{center}
  \includegraphics[width=0.5\linewidth]{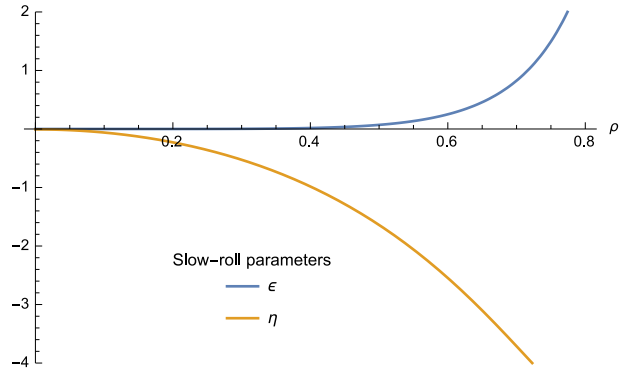}
 \caption{\small This plot shows the slow-roll parameters for  $F =0$, $b=3/2$, $\xi = -1.005$ and $q = 4.544 \times 10^{-7}$.}
 \label{slow_roll_AB_zero}
\end{center}
\end{figure*}

\begin{figure*}
\begin{center}
  \includegraphics[width=0.5\linewidth]{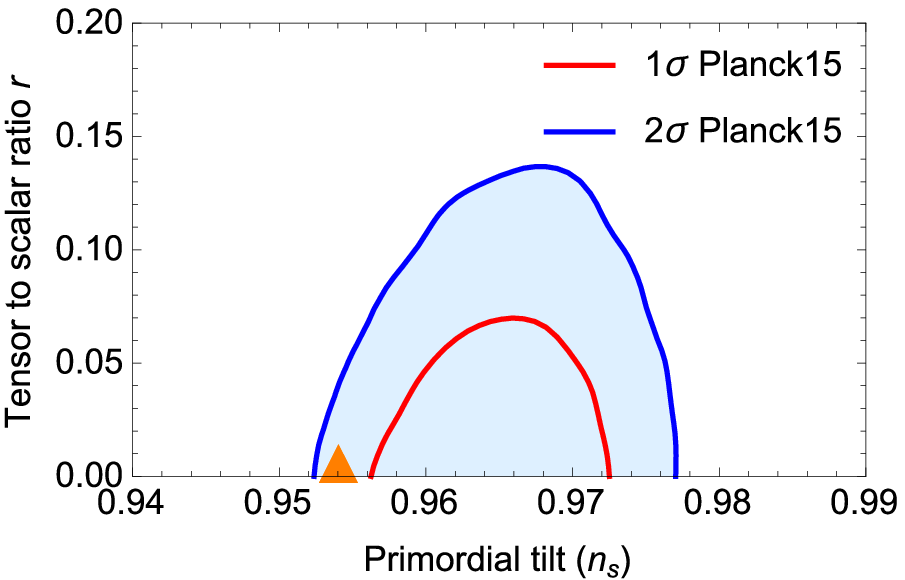}
 \caption{\small A plot of the predictions for the scalar potential with  $F =0$, $b=3/2$, $\xi = -1.005$ and $q = 4.544 \times 10^{-7}$ and the initial condition $\rho_* = 0.055$ and $\rho_{end} = 0.403$ in the $n_s$ - $r$ plane, versus Planck'15 results.}
 \label{ns_r_plot_AB_zero}
\end{center}
\end{figure*}

As an example, let us consider the case where 
 \begin{align}
q = 4.544 \times 10^{-7}, \quad \xi = -1.005.
 \label{parameter_AB_zero}
 \end{align}
The scalar potential for this parameter set is plotted in Fig.~\ref{potential_AB_Zero}.  The slow-roll parameters during inflation are shown in Fig.~\ref{slow_roll_AB_zero}. 
By choosing the initial condition $\rho_* = 0.055$ and $\rho_{\rm end} = 0.403$, we obtain the results $N = 58$, $n_s = 0.9542$, $r = 7.06 \times 10^{-6}$ and $A_s =2.2 \times 10^{-9}$ as shown in Table \ref{prediction_AB_zero} which are within the $2\sigma$-region of Planck'15 data  as shown in Fig.~\ref{ns_r_plot_AB_zero}. 

\begin{table}
  \centering 
  \begin{tabular}{|c|c|c|c|}
\hline
$N$ & $n_s$ & $r$ & $A_s$  \\
\hline
 58 & 0.9541 & $7.06 \times 10^{-6} $ & $2.22  \times 10^{-9} $  \\
\hline
\end{tabular}
  \caption{\small The theoretical predictions for $\rho_{*} = 0.055$ and $\rho_{\rm end} = 0.403 $ and the parameters given in eqs. (\ref{parameter_AB_zero}). }
  \label{prediction_AB_zero}
\end{table}

As was shown in Section~\ref{F0}, this model has a supersymmetric minimum with zero cosmological constant because $F$ is chosen to be zero. One possible way to generate a non-zero cosmological constant at the minimum is to turn on the superpotential $W = \kappa^{-3} F \neq 0$, as mentioned in Section~\ref{Fnot0}. In this case, the scale of the cosmological constant is of order $\mathcal{O}(F^2)$. It would be interesting to find an inflationary model which has a minimum at a tiny tuneable vacuum energy with a supersymmetry breaking scale consistent with the low energy particle physics.

\subsection{A small field inflation model from supergravity with observable tensor-to-scalar ratio}

\begin{figure*}
\begin{center}
  \includegraphics[width=0.5\linewidth]{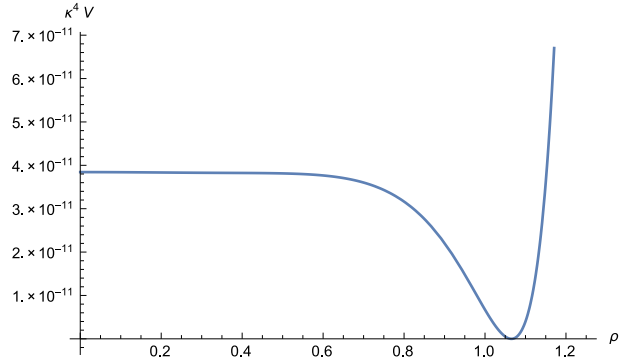}
 \caption{\small This plot shows the scalar potential for  $F =0$, $b=3/2$, $A = 0.545$, $B=0.230$, $\xi = -1.140$ and $q = 2.121 \times 10^{-5}$.}
 \label{potential_AB}
\end{center}
\end{figure*}

\begin{figure*}
\begin{center}
  \includegraphics[width=0.5\linewidth]{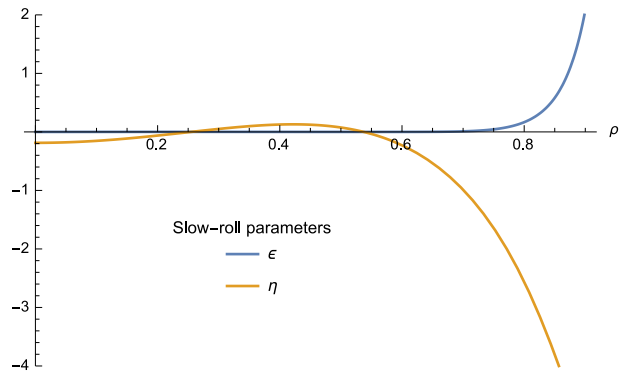}
 \caption{\small This plot shows the slow-roll parameters for  $F =0$, $b=3/2$, $A = 0.545$, $B=0.230$, $\xi = -1.140$ and $q = 2.121 \times 10^{-5}$.}
 \label{slow_roll_AB}
\end{center}
\end{figure*}

\begin{figure*}
\begin{center}
  \includegraphics[width=0.5\linewidth]{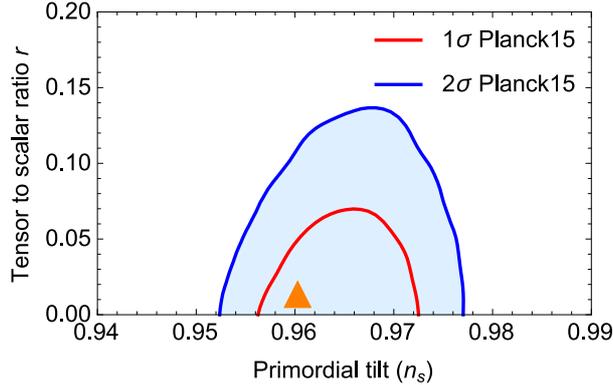}
 \caption{\small A plot of the predictions for the scalar potential with $F =0$, $b=3/2$, $A = 0.545$, $B=0.230$, $\xi = -1.140$ and $q = 2.121 \times 10^{-5}$ in the $n_s$ - $r$ plane, versus Planck'15 results.}
 \label{ns_r_plot_AB}
\end{center}
\end{figure*}

While the results in the previous example agree with the current limits on $r$ set by Planck, supergravity models with higher $r$ are of particular interest.  In this section we show  that our model can get large $r$ at the price of introducing some additional terms in the K\"ahler potential.  Let us consider the previous model with additional quadratic and cubic terms in $X\bar X$:
\begin{equation}
K = \kappa^{-2} \big(X\bar{X} + A (X\bar{X})^2 + B (X\bar{X})^3 + b \ln X\bar{X}\big),
\end{equation}
while the superpotential and the gauge kinetic function remain as in eq.~(\ref{const_W}).  We now assume that inflation is driven by the D-term, setting the parameter $F = 0$.  In terms of the field variable $\rho$, we obtain the scalar potential:
\begin{equation}
\mathcal{V} = q^2 \left( b+\rho^2 + 2A \rho^4 + 3B \rho^6  + \xi  \rho^{\frac{4 b}{3}} e^{\frac{2}{3} \left(A \rho ^4+B \rho ^6+\rho ^2\right)}\right)^2.
\end{equation}
We now have two more parameters $A$ and $B$.  This does not affect the arguments of the choices of $b$ in the previous sections because these parameters appear in higher orders in $\rho$ in the scalar potential.  So, we consider the case $b = {3}/{2}$.  The simple formula \eqref{simpleNe} for the number of e-folds for small $\rho^2$ also holds even when $A,B$ are turned on because the new parameters appear at order $\rho^4$ and higher. To obtain $r \approx 0.01$, we can choose for example 
 \begin{align}
 q = 2.121 \times 10^{-5}, \quad 
 \xi = -1.140, \quad
 A = 0.545, \quad
 B = 0.230.
 \label{parameter_AB}
 \end{align}
The scalar potential for these parameters is plotted in Fig.~\ref{potential_AB}.  The slow-roll parameters during inflation are shown in Fig.~\ref{slow_roll_AB}.  By choosing the initial condition $\rho_* = 0.240$ and $\rho_{\rm end} = 0.720$, we obtain the results $N = 57$, $n_s = 0.9603$, $r = 0.015$ and $A_s =2.2 \times 10^{-9}$ as shown in Table \ref{prediction_AB}, which agree with Planck'15 data as shown in Fig.~\ref{ns_r_plot_AB}.  

\begin{table}
  \centering 
  \begin{tabular}{|c|c|c|c|}
\hline
$N$ & $n_s$ & $r$ & $A_s$  \\
\hline
 57 & 0.9603 & $ 0.015$ & $2.22  \times 10^{-9} $  \\
\hline
\end{tabular}
  \caption{\small The theoretical predictions for $\rho_{*} = 0.055$ and $\rho_{\rm end} = 0.403 $ and the parameters given in Figure~\ref{potential_AB}. }
  \label{prediction_AB}
\end{table}

In summary, in contrast to the model in \cite{previouspaper} where the F-term contribution is dominant during inflation, here inflation is driven purely by a D-term. Moreover, a canonical K\"ahler potential (\ref{Kahlerpotential2}) together with two FI-parameters ($q$ and $\xi$) is enough to satisfy Planck'15 constraints, and no higher order correction to the K\"ahler potential is needed. However, to obtain a larger tensor-to-scalar ratio, we need to introduce perturbative corrections to the K\"ahler potential up to cubic order in $X\bar{X}$ (i.e. up to order $\rho^6$). This model provides a supersymmetric extension of the model \cite{Brustein18}, which realises large $r$ at small field inflation without referring to supersymmetry.

\section{Conclusions}

In this paper, we have shown that charged matter fields can be consistently coupled with the recently proposed FI-term \cite{Cribiori:2017laj} in the frame where the superpotential is invariant under the $U(1)$ symmetry.  We demonstrated that K\"ahler transformations do not give equivalent theories. It would be interesting to explore the possibility of recovering K\"ahler covariance but  obtaining the same physical action~\cite{nextproject}.

We then explored the possibility of obtaining inflation models driven by a D-term in the presence of the two FI terms. We first constrained one of the FI parameters by requiring that a slow-roll small-field inflation starts around the origin of the scalar potential which should be a local maximum. In the case where the superpotential vanishes, the potential has a global minimum preserving supersymmetry.  We found explicit models in which the slow-roll conditions are satisfied and inflation is driven by the D-term. Although the predicted tensor-to-scalar ratio of primordial perturbations is quite small for canonical K\"ahler potential, we found that by adding perturbative corrections, we can achieve significantly larger ratios that could be observed in the near future.

These models provide an alternative realisation of inflation driven by supersymmetry breaking identifying the inflaton with the goldstino superpartner~\cite{previouspaper}, but based on a D-term instead of an F-term.

We also discussed the case where the superpotential is turned on. Then, supersymmetry is broken at the global minimum but the supersymmetry breaking scale is of the order of the cosmological constant. In order to connect our model with low energy particle physics, one needs to find a mechanism for reconciling the hierarchy between the two scales in our model. 

\section*{Acknowledgements}
This work was supported in part by the Swiss National Science Foundation, in part by a CNRS PICS grant and in part by the ``CUniverse'' research promotion project by Chulalongkorn University (grant reference CUAASC).
The authors would like to thank Ramy Brustein, Toshifumi Noumi, Qaiser Shafi, and Antoine Van Proeyen for fruitful discussions.

\end{document}